\documentclass{aa}
\usepackage{graphicx}
\usepackage{natbib}
\usepackage{txfonts}
\bibpunct{(}{)}{;}{a}{}{,} 

\def\ls{LS~5039}
\def\lsi{LSI~+61$\degr$303}
\def\psrb{PSR~B1259-63}

\def\tfrac#1#2{{\textstyle {#1 \over #2}}}%

\begin{document}

\title{Spectral signature of a free pulsar wind in the gamma-ray
binaries LS~5039 and LSI~+61$\degr$303}
\titlerunning{Spectral signature of a free pulsar wind in LS~5039
and LSI~+61$\degr$303}

\author{
Benoit Cerutti 
\and Guillaume Dubus 
\and Gilles Henri 
}
\authorrunning{Cerutti, Dubus and Henri}
\institute{
Laboratoire d'Astrophysique de Grenoble, UMR 5571 CNRS, Universit\'e
Joseph Fourier, BP 53, 38041 Grenoble, France 
}

\date{Draft \today}
\abstract
{\ls\ and \lsi\ are two binaries that have been detected in  the TeV
energy domain.
These binaries are composed of a massive star and a compact object,
possibly a young pulsar. The gamma-ray emission would be due to
particle acceleration at the collision site between the relativistic
pulsar wind and the stellar wind of the massive star. Part of the
emission may also originate from inverse Compton scattering of stellar
photons on the unshocked (free) pulsar wind.}
{The purpose of this work is to constrain the bulk Lorentz
factor of the pulsar wind and the shock geometry in the compact pulsar
wind nebula scenario for \ls\ and \lsi\ by computing the unshocked
wind emission and comparing it to observations.}
{Anisotropic inverse Compton losses equations are derived and applied
to the free pulsar wind in binaries. The unshocked wind spectra seen
by the observer are calculated taking into account the
$\gamma-\gamma$ absorption and the shock geometry.}
{A pulsar wind composed of monoenergetic pairs produces a typical sharp peak at an energy which depends
on the bulk Lorentz factor and whose amplitude depends on the size of
the emitting region. This emission from the free pulsar wind is found to be strong and difficult to avoid in \ls\ and \lsi. }
{If the particles in the pulsar are monoenergetic then the observations constrain their energy to roughly 10-100 GeV. For more complex particle distributions, the free pulsar wind emission will be difficult to distinguish from the shocked pulsar wind emission.}

\maketitle

\section{Introduction}

Pulsars are fast rotating neutron stars that contain a large amount of
rotational energy. A significant fraction of this energy is carried away
by an ultra-relativistic wind of electrons/positrons pairs and possibly
ions (see \citealt{2007astro.ph..3116K} for a recent review). In the classical model of the Crab nebula
\citep{1974MNRAS.167....1R,1984ApJ...283..694K}, the pulsar wind is
isotropic, radial and monoenergetic with a bulk Lorentz factor
$\gamma_0\sim 10^6$ far from the light cylinder where the wind is
kinetic energy-dominated ($\sigma\ll1$). The cold relativistic wind
expands freely until the ram pressure is balanced by the surrounding
medium at the standoff distance $\rm{R_s}$. In the termination shock
region, the pairs are accelerated and their pitch angle to the
magnetic field are randomized, producing an intense synchrotron
source. Moreover, the inverse Compton scattering of the relativistic
electrons on soft photons produces high energy (HE, GeV domain) and very high
energy (VHE, TeV domain) gamma-rays.

The shocked pulsar wind is thought to be responsible for most of the
emitted radiation and gives clues about the properties of this
region. However, our knowledge of the unshocked pulsar wind region is
limited and based on theoretical statements. If the magnetic field is
frozen into the pair plasma as it is usually assumed, there is no
synchrotron radiation from the unshocked wind. Nevertheless, nothing
prevents inverse Compton scattering of soft photons onto the cold
ultra-relativistic pairs from occuring. The pulsar wind nebula (PWN)
emission has two components: radiation from the shocked and the
unshocked regions.

\citet{2000MNRAS.313..504B} investigated the inverse Compton emission
from the region upstream the termination shock of the Crab
pulsar. Comparisons between calculated and measured fluxes put limits
on the parameters of the wind, in particular the size of the kinetic
energy dominated region. \citet{2000APh....12..335B} investigated
emission from an unshocked freely expanding wind with no termination
shock in compact binaries. They computed spectra and light curves in
the gamma-ray binary \psrb, a system with a 48 ms pulsar and a
Be star in a highly eccentric orbit. The resulting gamma-ray emission is
a line-like spectrum.

In addition to \psrb, two other
binaries have been firmly confirmed as gamma-ray sources: \ls\
\citep{2005Sci...309..746A} and \lsi\
\citep{2006Sci...312.1771A}. They are composed of
a massive O or Be star and a compact object in an eccentric orbit. The
presence of a young pulsar was detected only in \psrb\
\citep{1992ApJ...387L..37J}. Radio pulses are
detectable but vanish near periastron, probably due to
free-free absorption and interaction with the Be disk wind. The
compact PWN scenario is most probably at work in this system and
investigations were carried out to model high and very high energy
radiation \citep{1999APh....10...31K,2005MNRAS.356..711S,2007MNRAS.380..320K,2008MNRAS.tmp..347S}. In \ls\ and
\lsi\, the
nature of the compact object is still controversial but spectral and
temporal similarities with \psrb\
argue in favor of the compact pulsar wind nebula scenario
\citep{2006A&A...456..801D}. The VHE radiation would therefore
be produced by the interaction between the pulsar wind and
the stellar companion wind. The massive star provides a huge density
of seed photons for inverse Compton scattering with the
ultra-relativistic pairs from the pulsar wind. Because of the relative
position of the compact object, the companion star and the observer,
the Compton emission is modulated on the orbital period. The vicinity of a massive
star is an opportunity to probe the pulsar wind at small scales.

The component of the shocked pulsar wind was computed in
\citet{2008A&A...477..691D} for \ls\ and limits on the
electron distribution, the pulsar luminosity and the magnetic field at
the termination shock were derived. \citet{2008arXiv0801.3427S}
calculated the VHE emission in \ls\ as well, assuming a power law
injection spectrum for the pairs in the unshocked pulsar wind and pair
cascading. In this paper, we investigate the anisotropic
inverse Compton scattering of stellar photons on the unshocked pulsar
wind within the compact PWN scenario for \ls\ and \lsi. Because of their tight orbits, the photon density is higher than in the Crab pulsar and \psrb. A more intense gamma-ray signal from the unshocked pulsar wind is expected. The main purpose
of this work is to constrain the bulk Lorentz factor $\gamma_0$ of the
pairs and the shock geometry. The next section presents the method and
the main equations used in order to compute spectra in gamma-ray
binaries. Section 3 describes and shows the expected spectra for \ls\
and \lsi\ with different parameters. Section 4 discusses the spectral signature from the unshocked pulsar wind.

\section{Anisotropic Compton losses in $\gamma$-binaries}

\subsection{The cooling of pairs}

An electron of energy $\rm{E_e}=\gamma_e m_e c^2$ in a given soft
photon field of density $n_0$ ph $\rm{cm}^{-3}$ cools down through
inverse Compton scattering (here the term `electrons' refers
indifferently to electrons and positrons). The power lost by the
electron is given by \citep{1965PhRv..137.1306J,1970RvMP...42..237B}
\begin{equation}
-\frac{dE_e}{dt}=\int^{\epsilon_+}_{\epsilon_-}\left(\epsilon_1-\epsilon_0\right)n_0\frac{dN}{dtd\epsilon_1}d\epsilon_1
\label{loss}
\end{equation}
where $\epsilon_0$ is the incoming soft photon energy, $\epsilon_1$
the scattered photon energy and dN/dtd$\epsilon_1$ is the Compton
kernel. $\epsilon_{\pm}$ boundaries are fixed by the relativistic
kinematics of inverse Compton scattering. The cooling of the pairs
$\rm{e^+/e^-}$ depends on the angular distribution and spectrum of
the incoming photon field. In the simple case of a
monoenergetic and unidirectional beam of photons in the Thomson limit,
the calculation of the Compton energy loss per electron is
\begin{equation}
-\frac{dE_e}{dt}=\sigma_T c n_0 \epsilon_0\left(1-\beta\mu_0\right)\left[\left(1-\beta\mu_0\right)\gamma^2_e-1\right]
\label{loss2}
\end{equation}
where $\sigma_T$ is the Thomson cross section, $\mu_0=\cos\theta_0$
and $\theta_0$ the angle between the incoming photon and the direction
of the electron motion. This calculation is done using the Compton
kernel calculated by \citet{fargion}. In the Thomson limit, the
cooling of the electron follow a $\gamma_e^2$ power law and has a
strong angular dependance. In a more general way and for
$\gamma_e\gg1$, the power lost per electron is calculated with the
kernel derived in \citet{2008A&A...477..691D} Eq.~(A.6).

\subsection{Compton cooling of the free pulsar wind}

\begin{figure}[h]
\centering
\includegraphics[width=8cm]{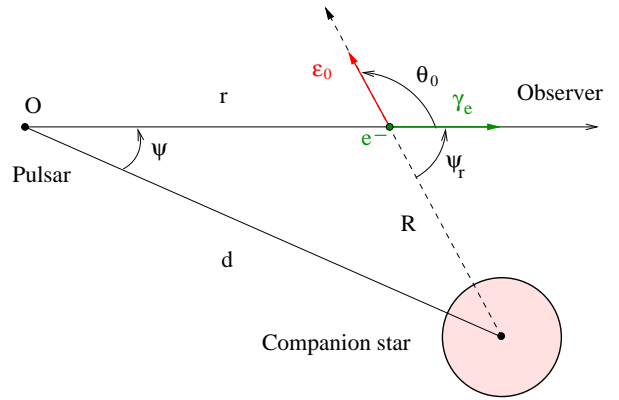}
   \caption{Geometry of the binary
     system. Electrons of Lorentz factor $\gamma_e$ are radially
     moving away at a distance r from the pulsar and R from the
     companion star. The angle $\psi$ quantifies the relative
     position between the pulsar, the companion star and the
     observer. $\psi_r$ measures the angle between the electron
     direction of motion and the line joining the companion star
     center to the electron position through its motion to the observer.}
    \label{geop}
\end{figure}

The pulsar is considered as a point-like source of monoenergetic and
radially expanding wind of relativistic pairs $e^+/e^-$. The pulsar
wind momentum is assumed to be entirely carried away by the pairs. The
companion star, with a typical luminosity of $10^{38}-10^{39}$ ergs
$\rm{s^{-1}}$, provides seeds photons for inverse
Compton scattering onto the radially expanding electrons from the pulsar. The
electrons see a highly anisotropic photon field. Inverse Compton
efficiency has a strong dependence on $\theta_0$ as
seen in Eq.~(\ref{loss2}). Depending of the relative position and
direction motion of the electron with respect to the incoming photons
direction, the cooling of the wind is anisotropic as well. Figure
\ref{geop} sketches the geometry considered in the binary system to
perform calculations.

\begin{figure}
\centering
\resizebox{\hsize}{!}{\includegraphics{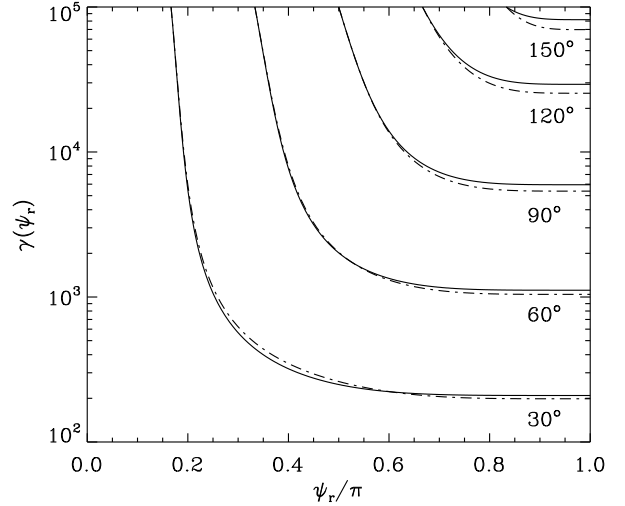}} 
   \caption{Compton cooling of a monoenergetic, free pulsar wind with
   $\gamma_0=10^5$, $d=2R_{\star}$ ($T_{\star}=39$ 000 K, 
   $R_{\star}=9.3$ $R_{\odot}$). The different curves show the
   dependence with the
   viewing angle $\psi$ on the cooling. $\psi$ varies between
   $30\degr$ (bottom) and
   $150\degr$ (top) if $i=60\degr$. Each curve shows the evolution
   of the Lorentz factor $\gamma$ with $\psi_r$ as the electron
   moves along the line of sight. $\psi_r$ is related to $r$ by
   Eq.~(\ref{rpsir}) so that $\psi_r=\psi$ at $r=0$ and $\psi_r=\pi$
   for $r=+\infty$. The calculation
   was carried out for a blackbody point like star (solid line)
   and taking into account the finite size of the star (dashed
   line).}
\label{ggam}
\end{figure}

For ultra-relativistic electrons , the radial dependence of the
electron Lorentz factor $\gamma_e(r)$ for a given viewing angle is
obtained by solving the first order differential equation
Eq.~(\ref{loss}).
\citet{1999MNRAS.304..359C} found an analytical solution in the
Thomson limit and \citet{2000APh....12..335B} derived a solution in
the general case using the \citet{1965PhRv..137.1306J} results for a
point-like and monoenergetic star with $\gamma_e\gg 1$. In this
approximation, the density of photons is $L_{\star}/(4\pi c
R^2\bar{\epsilon}_0)$ ph $\rm{cm^{-3}}$, where $L_{\star}$ is the star
luminosity and $\bar{\epsilon}_0=2.7kT_{\star}$ the average energy
photon from the star. The differential equation is then
\begin{equation}
\frac{d\gamma_e}{dr}=-\frac{1}{m_e c^3}\frac{L_{\star}}{4\pi c R^2}\int_{\epsilon_-}^{\epsilon_+}\left(\frac{\epsilon_1-\bar{\epsilon}_0}{\bar{\epsilon}_0}\right)\frac{dN}{dtd\epsilon_1}d\epsilon_1
\label{eqdiff2}
\end{equation}
where $R^2=d^2+r^2-2rd\cos\psi$. Calculations beyond the monoenergetic
and point-like star approximation require two extra integrations, one
over the star spectrum and the other onto the angular distribution of
the incoming photons due to the finite size of the star. The complete
differential equation is then given by
\begin{equation}
\frac{d\gamma_e}{dr}=-\frac{1}{m_e c^3}\iiint\left(\epsilon_1-\epsilon_0\right)n_0\frac{dN}{dtd\epsilon_1}d\epsilon_1 d\epsilon_0 d\Omega_0.
\label{eqdiff3}
\end{equation}
For a blackbody of temperature $T_{\star}$ and a spherical star of
radius $R_{\star}$, the incoming photon density $n_0$ is given by
Eq. (13) in \citet{2008A&A...477..691D}.
It is more
convenient to compute the calculation of the Lorentz factor as a
function of $\psi_r$ rather than $r$ (see Fig.~\ref{geop}). These two
variables are related through the relation
\begin{equation}
r=d\cos\psi\left(1-\frac{\tan\psi}{\tan\psi_r}\right),\hspace{2mm}r\in[0,+\infty],\hspace{2mm}\psi_r\in[\psi,\pi].
\label{rpsir}
\end{equation}
Figure \ref{ggam} presents the numerical computed output solution
$\gamma(\psi_r)$ applied to \ls\ with an inclination of $i=60\degr$
for a neutron star where the viewing angle varies between
$\pi/2-i=30\degr$ and $\pi/2+i=150\degr$. Here, the wind is assumed to
have an injection Lorentz factor $\gamma(\psi_r(0))=\gamma_0=10^5$ and
to continue unimpeded to infinity (i.e. it is not
contained by the stellar wind).

For small viewing angles $\psi$, the cooling of the wind is very efficient
because the collision electron/photon is almost head-on and the
electrons are moving in the direction of the star where the photon
density increases. For viewing angles $\psi\ga\pi/2$,
the cooling of the pairs is limited. In all cases, most of the cooling
occurs at
$\psi_r\sim\psi$. For $\psi_r\ga \pi/2$, the electron is moving away
from the star and the scattering angle become small leading to a
decrease in the wind energy loss. A comparison of Compton cooling between the
point-like and finite size star is shown in Fig.~\ref{ggam}. The
effects of the finite size of the star are significant in two
cases. The impact of the finite size of the star is important if the observer
is within the cone defined by the star and the electron at apex (see
\citealt{2008A&A...477..691D} for more details). For viewing angles
$\psi\la \arcsin(R_{\star}/d)$, the cooling is less efficient whereas
for $\psi\ga \pi-\arcsin(R_{\star}/d)$ it is more efficient as it can
be seen in the two extreme value of $\psi$ in figure \ref{ggam}. The
other situation occurs when the electrons travel close to the
companion star surface, for $\psi\la\pi/2$ and $\psi_r\ga\pi/2$. In
that case the angular distribution of the stellar photons is broad and
close head-on scatterings are possible, leading to more efficient cooling
compared with a point-like star. Nevertheless, these effects remain
small for \ls\ and \lsi\ and will be neglected in the following
spectral calculations.
\begin{figure*}
\resizebox{17cm}{!}
{\includegraphics{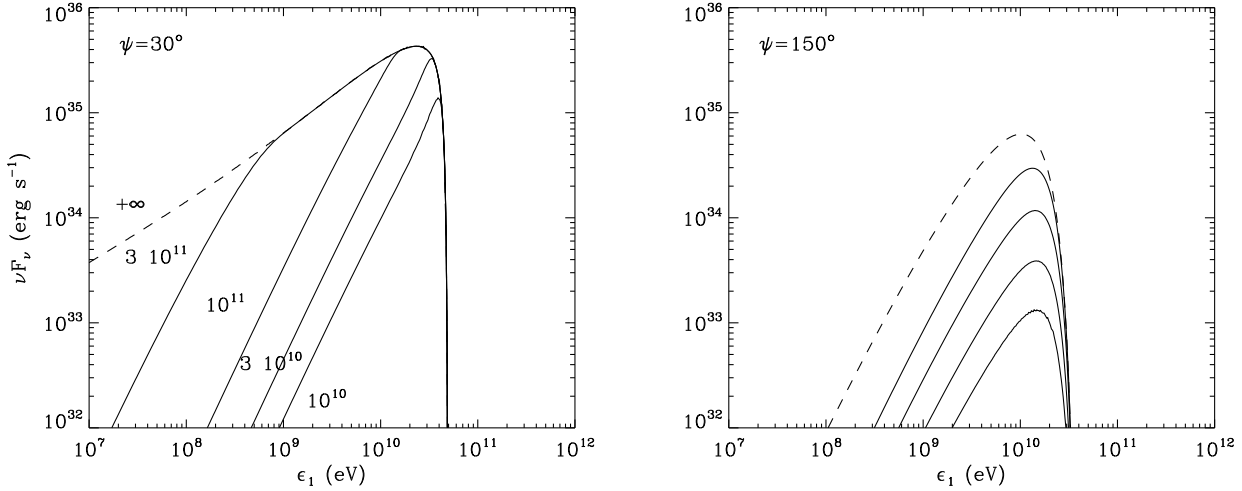}} 
   \caption{Computed inverse Compton spectrum from the unshocked pulsar wind
   in \ls\ and its dependence with the emitting region size
   $R_s$. The pulsar wind has  
   $\gamma_0=10^5$, $L_p=10^{36}$ erg $\rm{s^{-1}}$ and the star
   is a point-like blackbody. Spectra are calculated at the
   superior (left) and inferior (right) conjunctions for different
   standoff distances $R_s=10^{10}$ (bottom), $3$ $10^{10}$,
   $10^{11}$, $3$ $10^{11}$ cm and $+\infty$ (dashed line).}
\label{ic_spectre_fig}
\end{figure*}

\subsection{Unshocked pulsar wind spectra}

The number of scattered photons per unit of time, energy and solid
angle depends on three contributions: the density of the incoming
photons, the density of target electrons and the number of scattered
photons per electron. The pulsar wind of luminosity $L_p$ is
assumed isotropic and monoenergetic, composed only of pairs and with a
negligible magnetic energy density ($\sigma\ll1$). The
electrons density ($\rm{e}^-$ $\rm{cm}^{-3}$ $\rm{erg}^{-1}$) is then
proportional to $1/r^2$ if pair production is
neglected. Here, the interesting quantity for spectral calculations
is the number of electrons per unit of solid angle, energy and
length, which is $r^2$ time the electrons density so that
\citep{2000APh....12..335B} 
\begin{equation}
\frac{dN_e}{d\Omega_e d\gamma dr}=\frac{L_p}{4\pi c \beta_0 \gamma_0 m_e c^2}\delta\left(\gamma-\gamma_e(r)\right),
\label{edens}
\end{equation}
with $\delta$ the Dirac distribution. In deep Klein-Nishina regime, spectral broadening is expected because the continuous energy loss prescription fails ($\Delta E_e\sim E_e$). The complete kinetic equation must be used in order to describe accurately the electrons dynamics (see \citealt{1970RvMP...42..237B} Eq.~(5.7)). However, the $\delta$ approximation used here is reasonably good \citep{1989ApJ...342.1108Z}. The case of an anisotropic pulsar wind is discussed in \S4. 
In the following, the pulsar wind will therefore be assumed to follow
Eq.~(\ref{edens}). Heating of the pulsar wind by the radiative drag is
neglected \citep{2000APh....12..335B}.

In order to compute spectra, the emitted photons are supposed to be
entirely scattered in the direction of the electron motion. Because of
the ultra-relativistic motion of the electrons, most of the emission is
within a cone of aperture angle of the order of $1/\gamma_e\ll1$. In
this classical approximation, the spectrum seen by the observer is the
superposition of the contributions from the electrons along the line
of sight pulsar-observer in the solid angle $d\Omega_e$. The spectrum
seen by the observer is obtained with the following formula
\begin{equation}
\frac{dN_{tot}}{dt d\epsilon_1 d\Omega_e}=\iiiint n_0 \frac{dN}{dtd\epsilon_1}e^{-\tau_{\gamma\gamma}}\frac{dN_e}{d\Omega_e d\gamma dr}d\gamma d\epsilon_0 dr d\Omega_0
\label{spectrum}
\end{equation}
where $\tau_{\gamma\gamma}$ takes into account the absorption of
gamma-rays due to pair production with soft photons from the
companion star and is calculated following
\citet{2006A&A...451....9D}.

\subsection{The compact PWN geometry}

\begin{figure}[h]
\centering
\includegraphics[width=8cm]{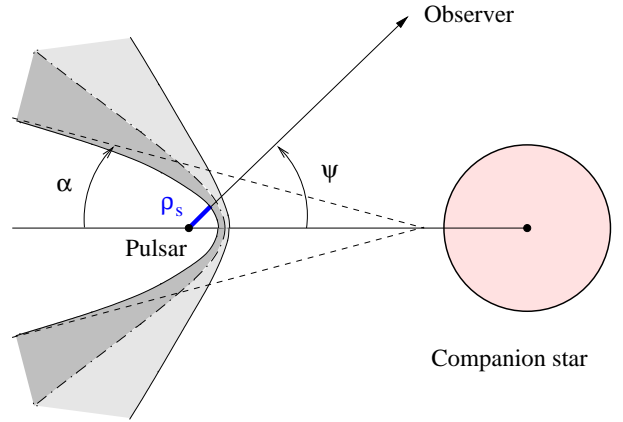}
   \caption{Shock geometry considered for the wind collision. For
     $\eta>1.25$ $10^{-2}$, the pulsar wind region
     remains open with an asymptotic half-opening
   angle $\alpha$. The dark region is the shocked relativistic
   pulsar wind and the light region is the shocked non-relativistic
   stellar wind, separated by a contact discontinuity (dot-dashed
   line). The size of the emitting zone seen by the observer
   $\rho_s$ depends on the viewing angle $\psi$.}
    \label{shock}
\end{figure}

The collision of the relativistic wind from the pulsar and the
non-relativistic wind from the massive star produces two termination
shock regions separated by a contact discontinuity (see
Fig. \ref{shock}). The geometry of the shock fronts are governed by
the ratio of the flux wind momentum quantified by $\eta$ and defined
as (e.g. \citealt{1992ApJ...386..265S,1993ApJ...402..271E})
\begin{equation}
\eta=\frac{L_p}{c \dot{M}_w v_{\infty}}
\label{eta}
\end{equation}
where $\dot{M}_w$ is the mass loss rate and $v_{\infty}$ the
stellar wind speed of the O/Be star. For two spherical winds, the
standoff distance point $R_s$ depends on $\eta$ and on the orbital
separation $d$
\begin{equation}
R_s=\frac{\sqrt{\eta}}{1+\sqrt{\eta}}d.
\label{rs}
\end{equation}
\citet{2008MNRAS.tmp..570B} have investigated the collision
between the pulsar wind and the stellar wind in the binary \psrb, with
a relativistic code and an isotropic pulsar wind in the hydrodynamical
limit. They obtained the geometry for the relativistic
and nonrelativistic shock fronts and the contact
discontinuity. They find that the collision
between the two winds produces an unclosed pulsar wind termination
shock (in the backward facing direction) for $\eta>1.25 \hspace{1mm}
10^{-2}$.\\

The size of the emitting region depends on the shock geometry
and the viewing angle, which can therefore have a major impact on the
emitted spectra. \citet{2001PASA...18...98B} computed spectra
from the unshocked pulsar wind in \psrb\ for an hyperbolic shock front
terminated close to the pulsar. They found a decrease in the spectra
fluxes and a decrease in the light curve asymmetry and flux
particularly near periastron compared with the spectra computed by
\citet{2000APh....12..335B}.

Figure \ref{ic_spectre_fig} presents computed spectra, ignoring
$\gamma\gamma$ absorption
at this stage, applied to \ls\ at the superior and inferior conjunctions
for different standoff distances $R_s$ and a pulsar wind with
$\gamma_0=10^5$. At the superior conjunction where $\psi=30\degr$, the
Compton cooling of the wind is efficient. The broadness in energy of
the radiated spectra is related to the size of the unshocked pulsar
wind region. For small
standoff distances $R_s\ll d$, spectra are truncated and sharp because
the termination shock region is very close to the pulsar, so that the pairs do not have time to radiate before reaching the shock. For $R_s\ga
d$, the free pulsar wind region is extended and emission from cooled
electrons starts 
contributing to the low energy tail in the scattered spectrum. The
amplitude of the spectrum reaches a maximum when the injected
particles can cool efficiently before reaching the shock. The spectral
luminosity is then set by the injected power and is not affected
anymore by the size of
the emitting zone. At the inferior conjunction where $\psi=150\degr$,
the cooling is less efficient and most of the emission occurs close to
the pulsar where the photon density and $\theta_0$ are greater,
regardless of the size of the emitting region. 
The radiated flux then depends linearly on $R_s$. A complete
investigation is presented in the next section where absorption and
spectra along the orbit are computed and applied to \ls\ and \lsi, ignoring pair cascading.

\section{Spectral signature of a monoenergetic pulsar wind in LS~5039 and LSI~+61$\degr$303}

In the following
sections, the emission expected by the unshocked pulsar wind in \ls\
and \lsi\ is compared with measured fluxes. Because spectra depends
on the shock geometry and the injection Lorentz factor, spectra are
calculated for various values of the two free parameters $\eta$ and $\gamma_0$.

\subsection{LS~5039}

\begin{figure}
\centering
\resizebox{\hsize}{!}{\includegraphics{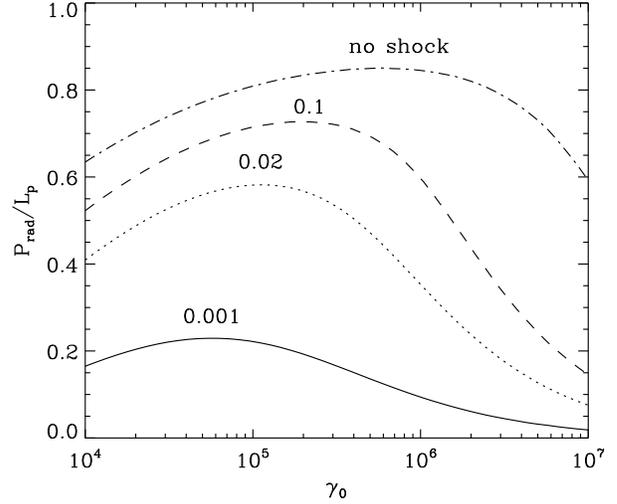}} 
   \caption{Total radiated power by the unshocked pulsar wind
     $P_{rad}$ in \ls\ as a function of $\gamma_0$. $P_{rad}$ is
     computed at periastron for $\eta=10^{-3}$ (solid line), 2
     $10^{-2}$ (dotted line), $10^{-1}$ (dashed line) and with no
     termination shock (dotted-dashed line).}
\label{frac_ls}
\end{figure}
\begin{figure*}
\centering
\resizebox{16cm}{!}
{\includegraphics{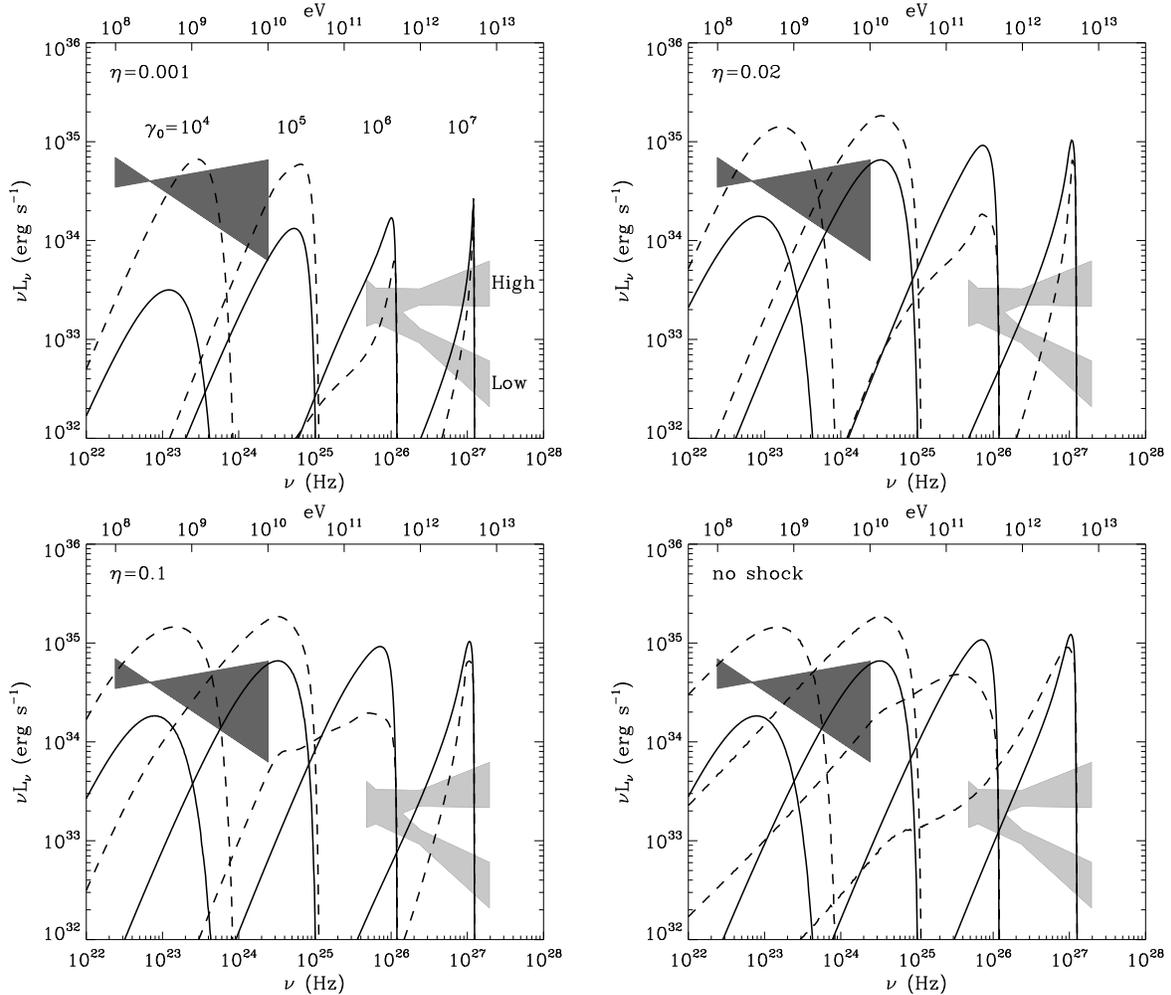}} 
   \caption{Spectral signature from the unshocked pulsar wind
     expected in LS 5039 and dependence with $\gamma_0$ and
     $\eta$. Spectra are averaged on the orbital phases
     corresponding to the HESS `high state' (solid
     line, $0.45<\phi<0.9$, with $\phi\equiv 0$ at periastron) and
     `low state' (dashed line,
     $\phi<0.45$ or $\phi>0.9$). The spectra are compared with
     EGRET (dark bowtie) and HESS (light bowties) observations, adopting a
     distance of 2.5 kpc. In the top left panel, $\eta=10^{-3}$ the
     shock is closed and the unshocked pulsar wind is assumed
     spherical.
     For $\eta=2$ $10^{-2}$
     (top right panel) and
     $\eta=0.1$ (bottom left panel) the shock is open
     with half-opening angles $\alpha\sim 2\degr$ and
     $\alpha\sim 30\degr$ respectively. The bottom right panel shows
     the extreme case with no
     termination shock.}
\label{ls_fig}
\end{figure*}

The companion star and the pulsar winds are assumed isotropic and
purely radial. The orbital parameters are those measured by
\citet{2005MNRAS.364..899C} as used in \citet{2008A&A...477..691D}.

Figure \ref{frac_ls} gives the total power radiated by the electrons in the
unshocked pulsar wind as a function of $\gamma_0$ at periastron. Here, the
shock front is assumed spherical of radius $R_s$. A maximum of
efficiency is observed at about $\gamma_0\sim 10^5$ which corresponds
to the transition between the Thomson and Klein-Nishina regimes where
the Compton timescale is shortest \citep{2006A&A...456..801D}. The
fraction of the pulsar wind power radiated at periastron depends
strongly on $R_s$. It is about $20\%$ for $\eta=10^{-3}$ and can
reach $70\%$ for $\eta=0.1$. Hence, most of the spindown energy
can be radiated directly by the unshocked pulsar wind.

Figure \ref{ls_fig} presents computed spectra averaged along the
orbit for different shock geometry and Lorentz factor with a pulsar
spindown luminosity of $L_p=10^{36}$ erg $\rm{s^{-1}}$. The
relativistic shock front is described by an hyperbolic equation. The
hyperbola apex is set by Eq.~(\ref{rs}) and the asymptotic
half-opening angle $\alpha$ is taken from Eq.~(27) in
\citet{2008MNRAS.tmp..570B},
both parameters depending only on $\eta$. Figure \ref{shock} sketches
the shock morphology for $1.25$ $10^{-2}<\eta<1$ and presents the
different shock fronts expected. The twist due to the orbital
motion is ignored since most of the emission occurs in the vicinity of the
pulsar. The size of the
emitting zone $\rho_s$ seen by the observer is thus set for any given
viewing angle $\psi$. Note that it is always greater than $R_s$. The
remaining free parameter $\gamma_0$ is chosen independently
between $10^4$ and $10^7$.

Computed spectra predict the presence of a narrow peak in the
spectral energy distribution due to the presence of the free pulsar
wind. The luminosity of this narrow peak can be comparable to or
greater than the measured fluxes by EGRET and HESS
\citep{1999ApJS..123...79H,2006A&A...460..743A}. For $\eta=10^{-3}$,
the pulsar wind termination shock is closed and the unshocked wind
emission zone is small. For $\eta=0.02$ and $\eta=0.1$ the line
spectra are well above both the limits imposed by the HESS
observations. The extreme case with no termination shock shows little
differences with the case where $\eta=0.1$.
Spectroscopic observations of \ls\ constrains
the O star wind parameters to  $\dot{M}_w\sim 10^{-7}$ $M_{\odot}$
$\rm{yr}^{-1}$ and $v_{\infty}\sim 2400$ km $\rm{s}^{-1}$
\citep{2004ApJ...600..927M}. Assuming $L_p=10^{36}$ erg $\rm{s}^{-1}$
then gives
$\eta\sim 2$ $10^{-2}$ (top right panel of Fig.~\ref{ls_fig}) or
$R_s\approx 2$ $10^{11}$~cm as in \citep{2006A&A...456..801D}. In this
case, almost  half of the pulsar wind energy is lost to inverse
Compton scattering before the shock is reached
(Fig.~\ref{frac_ls}). This is an upper limit since the reduced pulsar
wind luminosity would bring the shock location closer to the pulsar
than estimated from Eq.~(\ref{rs}). HESS observations already rule out
a monoenergetic pulsar wind with $\gamma_0=10^{6}$ or $10^7$ and 
$L_p=10^{36}$ erg $\rm{s^{-1}}$ as this would produce a large
component easily seen at all
orbital phases (see Fig.~5 in \citealt{2008A&A...477..691D}). The
EGRET observations probably also already rule out values of
$\gamma\leq 10^5$.

\subsection{LSI~+61$\degr$303}

In this system the stellar wind from the companion star is assumed to
be composed of a slow dense equatorial disk and a fast isotropic polar wind. 
The stellar wind may be clumpy and \citet{2008arXiv0802.1174Z} have
proposed a model of the high-energy emission from LSI~+61$\degr$303
that entails a mix between the stellar and the pulsar wind. The
orbital parameters are those measured by \citet{2005MNRAS.360.1105C}
(new orbital parameters were recently measured by
\citealt{2007ApJ...656..437G}).

Computed spectra applied to \lsi\ and averaged over the orbit to compare
with EGRET and MAGIC luminosities
\citep{1999ApJS..123...79H,2006Sci...312.1771A} are
presented in Fig. \ref{lsi_fig}. New data were recently reported by the MAGIC collaboration \citep{2008arXiv0806.1865M}. They confirmed the measurements of the first observational compaign and found a periodicity in the gamma-ray flux close to the orbital period. The pulsar spindown
luminosity is set to $L_p=10^{36}$ erg $\rm{s}^{-1}$ and the injected
Lorentz factor to $10^4$, $10^5$, $10^6$ and $10^7$ as for
\ls. There is more uncertainty in $\eta$ because of the complexity of
the stellar wind. The polar outflow is usually
modelled with $\dot{M}_w=10^{-8}$ $M_{\odot}$ $\rm{yr}^{-1}$ and
$v_{\infty}=2000$ km $\rm{s}^{-1}$ \citep{1988A&A...198..200W} leading
to $\eta\sim 0.2-0.3$. Concerning the slow dense equatorial disk, the
mass flux is typically one hundred times greater than the polar wind
and the terminal velocity is a few hundred km $s^{-1}$ giving $\eta$
compatible with $\sim 10^{-3}-10^{-2}$.

\begin{figure}
\centering
\resizebox{\hsize}{!}{\includegraphics{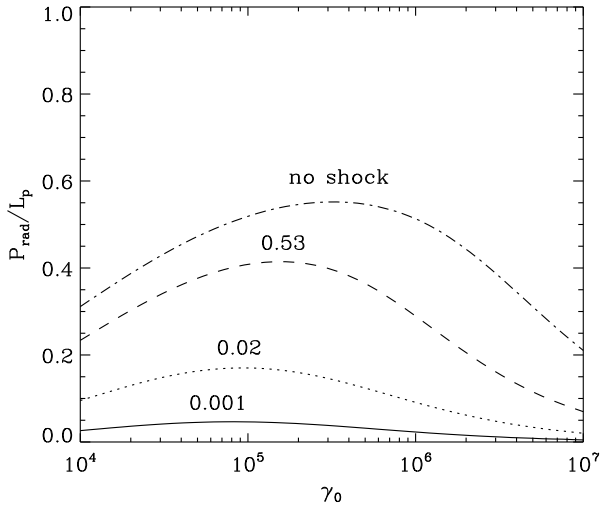}} 
   \caption{Total radiated power by the unshocked pulsar wind
     $P_{rad}$ in \lsi\ as a function of $\gamma_0$. $P_{rad}$ is
     computed at periastron for $\eta=10^{-3}$ (solid line), 2
     $10^{-2}$ (dotted line), 0.53 (dashed line) and with no
     termination shock (dotted-dashed line).}
\label{frac_lsi}
\end{figure}

\begin{figure*}
\centering
\resizebox{16cm}{!}
{\includegraphics{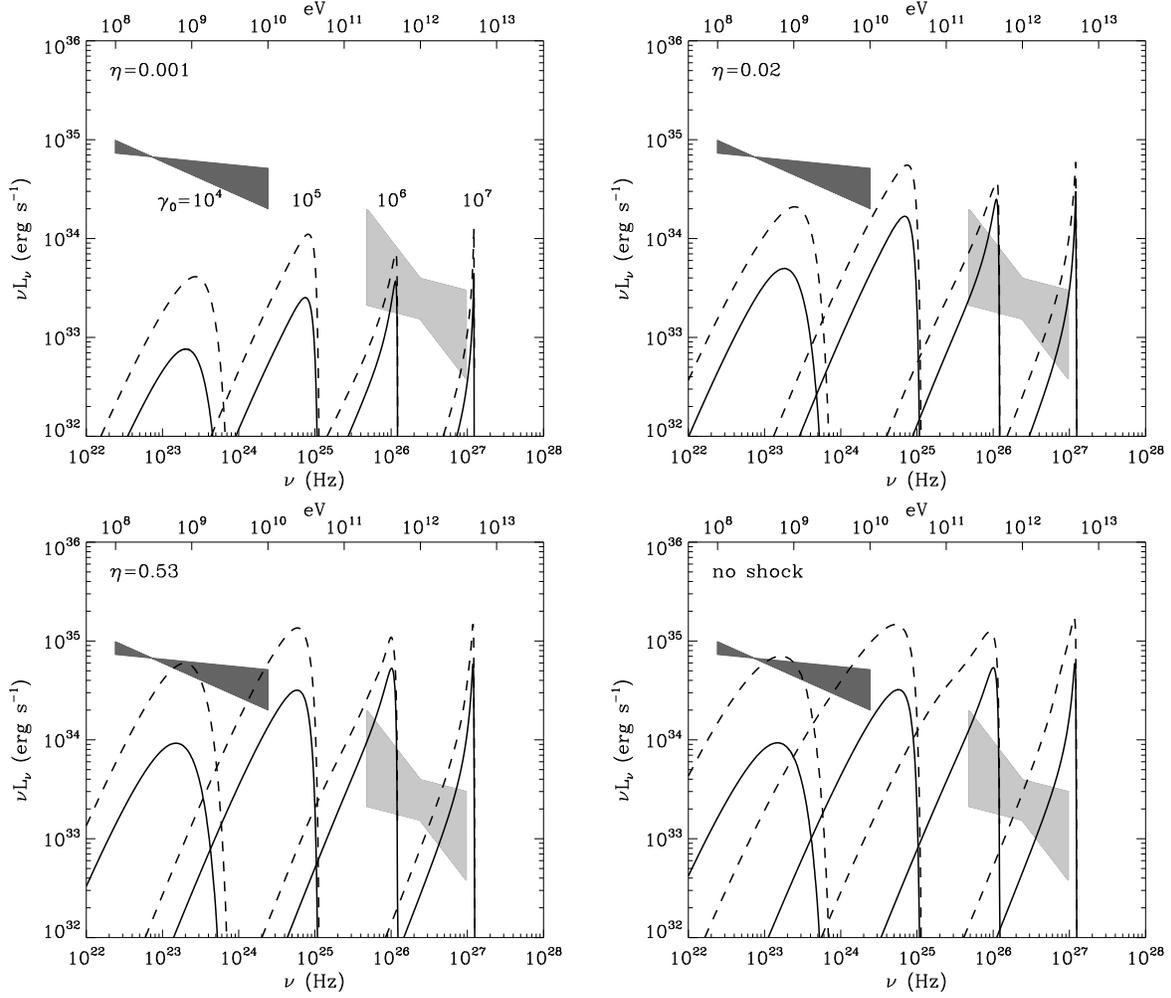}} 
   \caption{Spectral signature from the unshocked pulsar wind
     expected in \lsi\ and dependence with $\gamma_0$ and
     $\eta$. Spectra are averaged between phase $0.4<\phi<0.7$
     (solid line, at periastron $\phi\equiv 0.23$) and the complementary
     phases $\phi<0.4$ or $\phi>0.7$ (dashed
     line). Luminosities are compared with the EGRET (dark bowtie) and
     MAGIC (light bowtie) observations, adopting a
     distance of 2.3 kpc. In the top left panel, $\eta=10^{-3}$
     the shock is closed and the unshocked pulsar wind is assumed
     spherical. For $\eta=2$ $10^{-2}$ (top right panel) and
     $\eta=0.53$ (bottom left panel) the shock is open
     with half-opening angles $\alpha\sim 2\degr$ and
     $\alpha\sim 60\degr$ respectively. The bottom right panel shows
     the extreme case with no
     termination shock.}
\label{lsi_fig}
\end{figure*}

The overall behaviour is similar to \ls. The spectral luminosities
and the total power radiated by the unshocked pulsar wind (Fig.~\ref{frac_lsi})
are lower in \lsi\ than \ls\ because the compact object is more
distant to its companion star and the latter has a lower luminosity,
leading to a decrease in the density of seed photons for inverse
Compton scattering. If $\eta=10^{-3}$, no constrains on $\gamma_0$ can
be formulated as the spectrum is always below the observational
limits. For larger values of $\eta$, the very high energy observations
constrain $\gamma_0$ to below $10^{6}$, assuming the pulsar wind is
monoenergetic. Spectra were computed for $\eta=0.53$ with
$\dot{M}_w=10^{-8}$ $M_{\odot}$ $\rm{yr}^{-1}$ and $v_{\infty}=1000$
km $\rm{s}^{-1}$ as used by \citet{2007A&A...474...15R}. In this case,
the spectra are close to the freely propagating pulsar wind. The EGRET
luminosity is slightly overestimated for $\gamma_0\le 10^5$ when $\eta=0.53$.

\section{Discussion}

The proximity of the massive star in LS 5039 and \lsi\  provides an opportunity to directly probe the distribution of particles in the highly relativistic pulsar wind. The calculations show the inverse Compton emission from the unshocked wind should be a significant contributor to the observed spectrum. For a monoenergetic and isotropic pulsar wind the emission remains line-like, with some broadening due to cooling, as had been found previously for the Crab and PSR B1259-63 \citep{2000MNRAS.313..504B,2000APh....12..335B}. However, here, such line emission can pretty much be excluded by the available very high energy observations of HESS or MAGIC, and (to a lesser extent) by the EGRET observations that show power-law spectra at lower flux levels.

\subsection{Is the pulsar wind power overestimated?}

Reducing the pulsar power (or, equivalently, increasing the distance to the object) would diminish the predicted unshocked wind emission relative to the observed emission. This is not viable as this would also reduce the level of the shocked pulsar wind emission. Similarly, the energy carried by the particles may represent only a small fraction of the wind energy. At distances of order of the pulsar light cylinder the energy is mostly electromagnetic. Evidence that this energy is converted to the kinetic energy of the particles comes from plerions, which probe distances of order 0.1~pc from the pulsar. It is therefore conceivable that this conversion is not complete at the distances under consideration here (0.01-0.1 AU). In this case the emission from the particle component would be reduced. However, the shocked emission would also be reduced as high $\sigma$ shocks divert little of the energy into the particles \citep{1984ApJ...283..694K}. Furthermore, the high energy particles would preferentially emit synchrotron rather than inverse Compton due to the higher magnetic field. Hence,  this possibility also seems unlikely.

Alternatively, the unshocked wind emission could be weaker compared to the shocked wind emission if the termination shock was closer to the pulsar, {\em i.e.} if one had a low $\eta$. In \ls\ the unshocked wind emission is strong even with $\eta=0.001$, which already implies a stronger stellar wind than optical observations seem to warrant. Furthermore, the value of the magnetic field would be high if the termination shock was close to the pulsar and this inhibits the formation of very high energy gamma-rays as the high energy electrons would then preferentially lose energy to synchrotron radiation \citep{2006A&A...456..801D}. Hence, it does not seem viable either to invoke a smaller zone for the free wind. 

The conclusion is that the strong emission from the pulsar wind found in the previous section is robust against general changes in the parameters used. The following subsections examine how this emission can be made consistent with the observations.

\subsection{Constrains on the pulsar wind Lorentz factor}

The high level of unshocked emission is
compatible with the observations only if it occurs around 10 GeV  or above 10~TeV, {\em i.e.} outside of the ranges probed by EGRET and the current generation of Cherenkov telescopes.  This poses stringent constraints on the energy of the particles in the pulsar wind. The Lorentz factor of the pulsar wind would be constrained to a few $10^5$ or to more than $10^7$. The 1-100~GeV energy range will be partly probed by GLAST and HESS-2, and CTA in the more distant future. For instance, unshocked wind emission in \ls\ would appear in the GLAST data as a spectral hardening at the highest energies. Nevertheless, that the free wind is emitting in the least accessible spectral region may appear too fortuitous for comfort. 

\subsection{Anisotropic pulsar wind}

The assumptions on the pulsar wind may be inaccurate. Pulsar winds are thought to be
anisotropic \citep{1992ApJ...397..187B}. \citet{2002AstL...28..373B} interpreted the jet-torus structure revealed by X-ray Chandra observations of the Crab nebula, as a latitude dependence of the Lorentz factor $\gamma(\theta)=\gamma_i+\gamma_m \sin^2 \theta$ where $\gamma_i$ is small (say $10^4$) and $\gamma_m$ is high (say $10^6$). This hypothesis was corroborated by computational calculations in \citet{2004MNRAS.349..779K} where the synchrotron jet-torus was obtained.  Here, the pulsar orientation to the observer is fixed (unless it precesses) so that the initial Lorentz factor of the pulsar wind along the line of sight would remain the same along the orbit. However, assuming the particle flux in the pulsar wind remains isotropic, the unshocked wind emission will appear at a lower energy and at a lower flux if the pulsar is seen more pole-on. The peak energy of the line-like spectral feature directly depends on $\gamma(\theta)$. Its intensity will also decrease in proportion as the pulsar power matches the latitude change in $\gamma$ to keep the particle flux isotropic (see Eq.~6). 

The shocked wind emission is set by the mean power and Lorentz factor of the wind  and is insensitive to orientation. However, a more pole-on orientation will lower the contribution from the unshocked component. For instance, if  $\gamma(\theta)=10^4+10^6 \sin^2 \theta$ and $\theta=17\fdg5$ then the effective $\gamma$ along the line-of-sight will be $10^5$ and the observed luminosity of the unshocked emission will be lowered by a factor 10 compared to the mean  pulsar power (Eq. 6). The probability to have an orientation corresponding to a value of $\gamma(\theta)$ of 0.1 $\gamma_m$ or less is about $\tfrac{1}{5}$, assuming a uniform distribution of orientations. This would be enough to push the line emission to lower energies and to lower fluxes by a factor 10 or more, thereby relaxing the constraints on the mean Lorentz factor of the wind. Although this is not improbable, it would again require some fortuitous coincidence for the pulsars in both \ls\ and \lsi\ to be seen close enough to the pole that their free wind emission is not detected.


\subsection{The energy distribution of the pairs}

\begin{figure}
\centering
\resizebox{\hsize}{!}
{\includegraphics{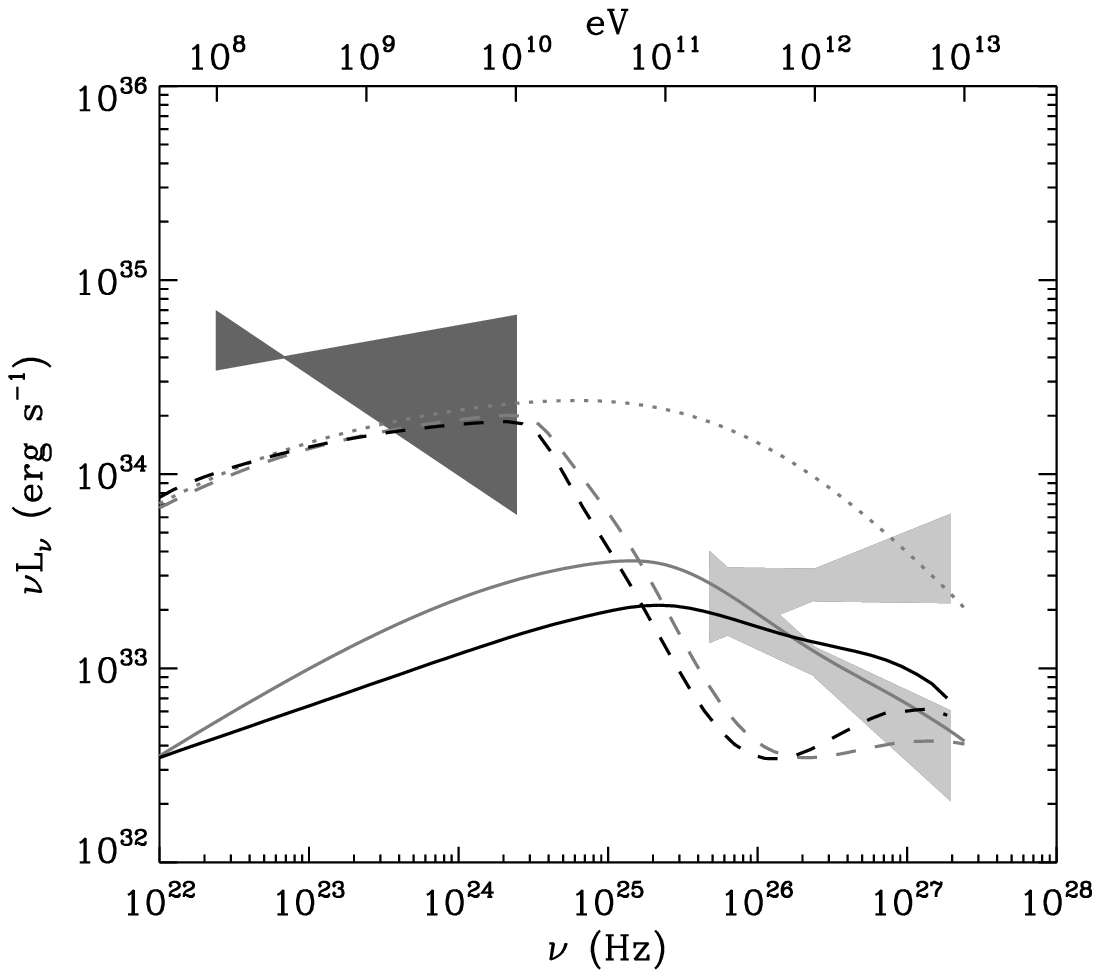}}
\resizebox{\hsize}{!}
{\includegraphics{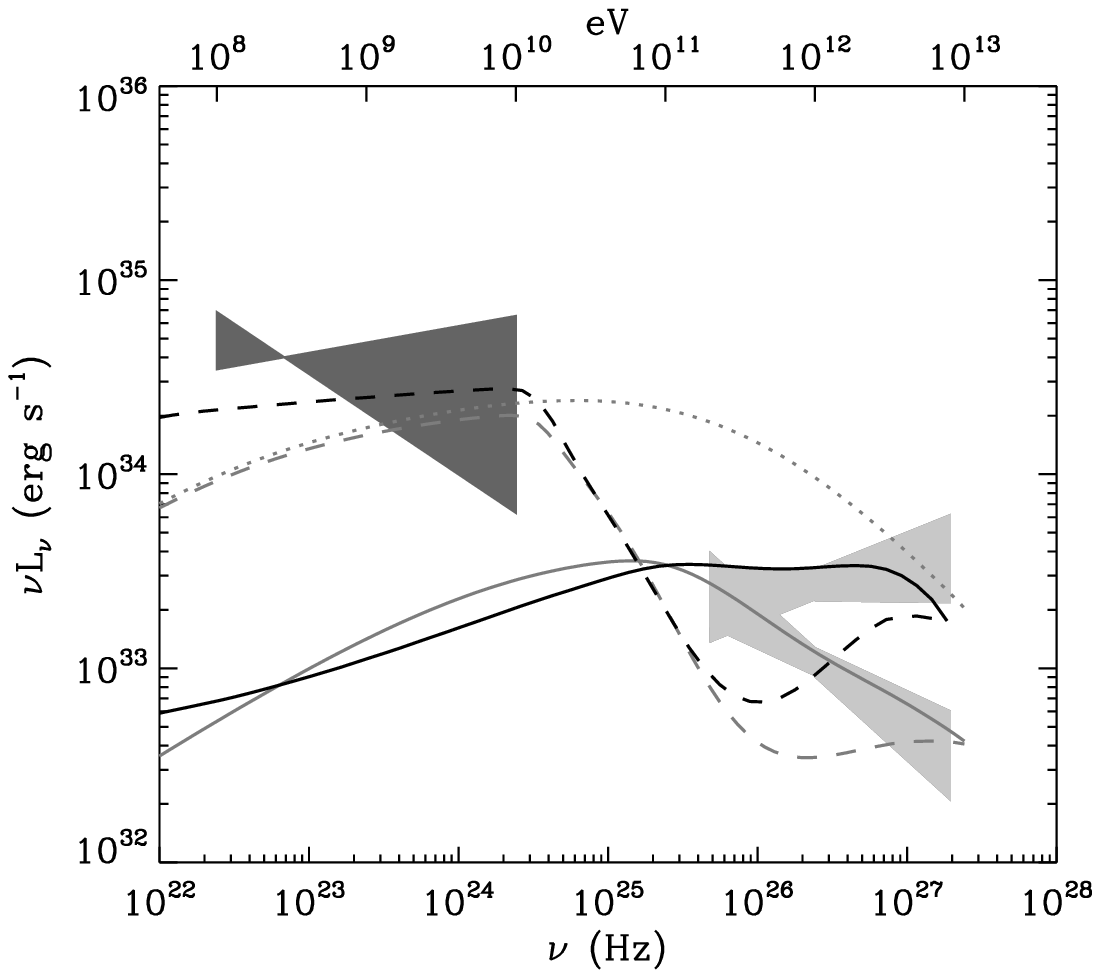}} 
   \caption{Comparison between emission from the shocked and unshocked regions in \ls, taking the same particle injection for both regions. The distribution is a power-law of index -2 between $\gamma=10^3$ and $10^8$ with total power $10^{36}$~erg~s$^{-1}$. Spectra are averaged to correspond to the HESS `high state' (solid lines) and `low state' (dashed lines) as in Fig.~\ref{ls_fig}. The geometry is a sphere of radius 2$~10^{11}$~cm. Grey lines show emission from the unshocked emission and dark lines show the emission from the shocked region. The orbital averaged {\em non-absorbed} spectrum from the unshocked pulsar wind is shown in grey dotted line. Particles escape from  the shocked region on a timescale $t_{\rm esc}=R_s/(c/3)$ (top) or $10 R_s/(c/3)$ (bottom). The unshocked emission is the same in both panels. }
\label{ls_pow1}
\end{figure}

The assumption of a monoenergetic wind may be incorrect, if only because the particles in the pulsar wind are bathed by a strong external photon field even as they accelerate and that this may lead to a significantly different distribution. Fig.~\ref{ls_pow1} shows the emission from a pulsar wind where the particles have been assumed to have a power-law distribution with an index of -2 between $\gamma$ of $10^3$ and $10^8$. Obviously, a power-law distribution of pairs erases the line-like spectral feature. The emission properties are essentially identical to the emission from the shocked region with a harder and fainter {\em intrinsic} Compton spectrum when the pulsar is seen in front of the star compared to when it is behind. The emission from the shocked region is also shown, calculated as in \citet{2008A&A...477..691D}. The particle injection spectrum is the same in both regions. The particles are assumed to stay close to the pulsar and to escape from the shocked region after a time $t_{\rm esc}=R_s/(c/3)$ (top) and $10 R_s/(c/3)$ (bottom). Longer $t_{\rm esc}$ do not change the distribution any further. The longer escape timescale enables a harder particle distribution to emerge at high energies (where the radiative timescale is comparable to $R_s/c$, see Fig.~2 in \citealt{2006A&A...456..801D}). With $t_{\rm esc}=R_s/(c/3)$, the shocked spectra is very  close to the unshocked spectra. Generally, calculations show the spectra from the shocked and unshocked regions may be indistiguishable when the injected particles are taken to be the same in both regions. The only possible difference is that the longer residence time of particles in the shocked region allows for harder spectra.

\subsection{Dominant emission from the unshocked wind}

\begin{figure}[t]
\centering
\resizebox{\hsize}{!}
{\includegraphics{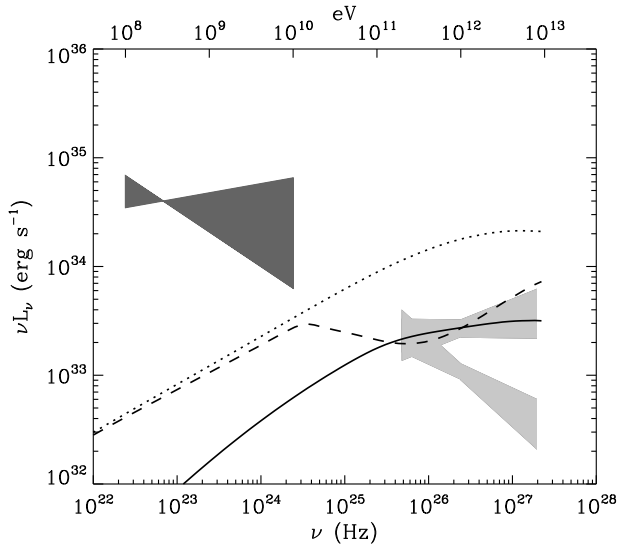}} 
   \caption{Fit to the observations of \ls, assuming the dominant contribution comes from the unshocked region of the pulsar wind. The `high state' of HESS corresponds to the solid line, the `low state' to the dashed line and the dotted line to the orbital averaged non-absorbed spectrum. The injected particles had a power law distribution with an index of $-1.5$ between $\gamma=10^3$ and 10$^8$. The injected power is 4 $10^{35}$ erg~s$^{-1}$.}
\label{ls_pow2}
\end{figure}

Emission from the unshocked wind could be the dominant contributor to the spectral energy distribution. In this case, the observations give the particle distribution in the pulsar wind.
\citet{2008arXiv0801.3427S}  have
considered such a scenario for \ls\ and use a total energy in leptons of about $10^{35}$ erg~s$^{-1}$ and a power-law distribution with an index around $-2$, both of which are adjusted to the observations and vary with orbital phase. The total pulsar power is much larger, 10$^{37}$~erg~s$^{-1}$, in order to have a big enough free wind emission zone. Most of the pulsar energy is then carried by nuclei. Such a large luminosity would make the pulsar very young, comparable to the Crab pulsar, implying a high birth rate for  such systems. Fig.~\ref{ls_pow2} shows the expected emission from a pulsar wind propagating to infinity and with a particle power law index of -1.5 from $\gamma=10^3$ to $10^8$ chosen to adjust the `high state' of LS 5039. The injected power is 4\ 10$^{35}$~erg~s$^{-1}$. The injected spectrum gives a good fit of the `hard' state. However, the `low' state dominates the complete very high energy contribution ($\ga$ 1~TeV) due to the extended emitting region. Particles have enough time to radiate very high energy gammay-ray far away, where they are less affected by $\gamma-\gamma$ absorption.

A possible drawback of this scenario is that  the synchrotron and inverse Compton emission are not tied by the shock conditions and that the total energy in leptons is low so that it is not clear how the radio, X-ray and gamma-ray observations below a few GeV would arise. It is also unclear how this can lead to a collimated radio outflow as seen in \lsi. A possibility is emission from secondary pairs created in the stellar wind by cascading as suggested by \citet{2008A&A...482..397B}. More work is necessary to understand these different contributions and the signatures that may enable to distinguish them. 

Another potential drawback of this scenario is that it does not explain why the very high energy gamma-ray flux is observed to peak close to apastron in \lsi. If the inverse Compton scattering in the pulsar wind is responsible then the maximum should be around periastron, especially as gamma-gamma attenuation is very limited in \lsi.  On the other hand, if the emission arises from the shocked pulsar wind then synchrotron losses explain the lack of very high energy gamma-rays at periastron: the pulsar probes the dense equatorial  wind from the Be star and the shock forms at a small distance, leading to a high magnetic field and a cutoff in the high-energy spectrum  (see \S6.2.2 in \citealt{2006A&A...456..801D}).

\section{Conclusion}

Gamma-ray binaries are of particular interest in the study of
pulsar wind nebula at very small scales. The massive star radiates a
large amount of soft seeds photons for inverse Compton scattering on
relativistic electrons from the pulsar. One expects two
contributions in the gamma-ray spectral energy distribution:
one from the shocked pulsar wind and another from the unshocked pulsar wind.
The spectral signature from the unshocked region is strong and depends
on the shock geometry and the initial energy of the pairs in the
pulsar wind. A significant fraction of the pulsar wind power can be
lost to inverse Compton scattering before a shock forms with the stellar wind \citep{2005MNRAS.356..711S}.
The shock location will thus be slightly closer in to the neutron star than calculated without taking into account the losses in the pulsar wind, assuming the wind is composed only of e$^+$e$^-$ pairs. 

Significant emission from the free pulsar wind seems unavoidable. Inverse Compton losses in the free wind may be reduced if the shock occurs very close to the neutron star. This is unlikely as it would require a very strong stellar wind or a pulsar wind power that would be too low to produce the observed emission. If the pulsar wind is anisotropic then the orientation of the pulsar with respect to the observer can make the unshocked emission less conspicuous. This comes at the price of a peculiar orientation. If the pulsar wind is monoenergetic, then the line-like expected spectrum exceeds the observed very high
energy power-laws for all geometries unless the pair energy is around 10 GeV or above 10 TeV. This
pushes the direct emission from the wind in ranges where it may be constrained by future GLAST, HESS-2 or CTA measurements. The absence of any line emission will show that the assumption of a Crab-like monoenergetic, low $\sigma$ pulsar wind was simplistic. If the pairs in the pulsar wind have a power-law distribution, then the unshocked emission is essentially indistinguishable from the shocked emission \citep{2008arXiv0801.3427S}. A promising alternative is the {\em striped wind} model in which the wind remains highly magnetised up to the termination shock, where the alternating field could be dissipated and accelerate particles (see \citealt{2007astro.ph..3116K} and references therein). Future theoretical studies on the generation of pulsar relativistic winds in the context of a strong source of photons may be able to yield the particle distribution to expect and lead to more accurate predictions.

\begin{acknowledgements}
GD acknowledges support from the {\em Agence Nationale de la
  Recherche}.
\end{acknowledgements}

\bibliographystyle{aa}
\bibliography{unsh}

\begin{thebibliography}{39}
\expandafter\ifx\csname natexlab\endcsname\relax\def\natexlab#1{#1}\fi

\bibitem[{{Aharonian} {et~al.}(2005){Aharonian}, {Akhperjanian}, {Aye},
  {Bazer-Bachi}, {Beilicke}, {Benbow}, {Berge}, {Berghaus}, {Bernl{\"o}hr},
  {Boisson}, {Bolz}, {Borrel}, {Braun}, {Breitling}, {Brown}, {Gordo},
  {Chadwick}, {Chounet}, {Cornils}, {Costamante}, {Degrange}, {Dickinson},
  {Djannati-Ata{\"i}}, {Drury}, {Dubus}, {Emmanoulopoulos}, {Espigat},
  {Feinstein}, {Fleury}, {Fontaine}, {Fuchs}, {Funk}, {Gallant}, {Giebels},
  {Gillessen}, {Glicenstein}, {Goret}, {Hadjichristidis}, {Hauser},
  {Heinzelmann}, {Henri}, {Hermann}, {Hinton}, {Hofmann}, {Holleran}, {Horns},
  {Jacholkowska}, {de Jager}, {Kh{\'e}lifi}, {Komin}, {Konopelko}, {Latham},
  {Le Gallou}, {Lemi{\`e}re}, {Lemoine-Goumard}, {Leroy}, {Lohse}, {Marcowith},
  {Martin}, {Martineau-Huynh}, {Masterson}, {McComb}, {de Naurois}, {Nolan},
  {Noutsos}, {Orford}, {Osborne}, {Ouchrif}, {Panter}, {Pelletier}, {Pita},
  {P{\"u}hlhofer}, {Punch}, {Raubenheimer}, {Raue}, {Raux}, {Rayner}, {Reimer},
  {Reimer}, {Ripken}, {Rob}, {Rolland}, {Rowell}, {Sahakian}, {Saug{\'e}},
  {Schlenker}, {Schlickeiser}, {Schuster}, {Schwanke}, {Siewert}, {Sol},
  {Spangler}, {Steenkamp}, {Stegmann}, {Tavernet}, {Terrier}, {Th{\'e}oret},
  {Tluczykont}, {Vasileiadis}, {Venter}, {Vincent}, {V{\"o}lk}, \&
  {Wagner}}]{2005Sci...309..746A}
{Aharonian}, F., {Akhperjanian}, A.~G., {Aye}, K.-M., {et~al.} 2005, Science,
  309, 746

\bibitem[{{Aharonian} {et~al.}(2006){Aharonian}, {Akhperjanian}, {Bazer-Bachi},
  {Beilicke}, {Benbow}, {Berge}, {Bernl{\"o}hr}, {Boisson}, {Bolz}, {Borrel},
  {Braun}, {Brown}, {B{\"u}hler}, {B{\"u}sching}, {Carrigan}, {Chadwick},
  {Chounet}, {Cornils}, {Costamante}, {Degrange}, {Dickinson},
  {Djannati-Ata{\"i}}, {O'C.~Drury}, {Dubus}, {Egberts}, {Emmanoulopoulos},
  {Espigat}, {Feinstein}, {Ferrero}, {Fiasson}, {Fontaine}, {Funk}, {Funk},
  {F{\"u}{\ss}ling}, {Gallant}, {Giebels}, {Glicenstein}, {Goret},
  {Hadjichristidis}, {Hauser}, {Hauser}, {Heinzelmann}, {Henri}, {Hermann},
  {Hinton}, {Hoffmann}, {Hofmann}, {Holleran}, {Horns}, {Jacholkowska}, {de
  Jager}, {Kendziorra}, {Kh{\'e}lifi}, {Komin}, {Konopelko}, {Kosack},
  {Latham}, {Le Gallou}, {Lemi{\`e}re}, {Lemoine-Goumard}, {Lohse}, {Martin},
  {Martineau-Huynh}, {Marcowith}, {Masterson}, {Maurin}, {McComb}, {Moulin},
  {de Naurois}, {Nedbal}, {Nolan}, {Noutsos}, {Orford}, {Osborne}, {Ouchrif},
  {Panter}, {Pelletier}, {Pita}, {P{\"u}hlhofer}, {Punch}, {Raubenheimer},
  {Raue}, {Rayner}, {Reimer}, {Reimer}, {Ripken}, {Rob}, {Rolland}, {Rowell},
  {Sahakian}, {Santangelo}, {Saug{\'e}}, {Schlenker}, {Schlickeiser},
  {Schr{\"o}der}, {Schwanke}, {Schwarzburg}, {Shalchi}, {Sol}, {Spangler},
  {Spanier}, {Steenkamp}, {Stegmann}, {Superina}, {Tavernet}, {Terrier},
  {Tluczykont}, {van Eldik}, {Vasileiadis}, {Venter}, {Vincent}, {V{\"o}lk},
  {Wagner}, \& {Ward}}]{2006A&A...460..743A}
{Aharonian}, F., {Akhperjanian}, A.~G., {Bazer-Bachi}, A.~R., {et~al.} 2006,
  \aap, 460, 743

\bibitem[{{Albert} {et~al.}(2006){Albert}, {Aliu}, {Anderhub}, {Antoranz},
  {Armada}, {Asensio}, {Baixeras}, {Barrio}, {Bartelt}, {Bartko}, {Bastieri},
  {Bavikadi}, {Bednarek}, {Berger}, {Bigongiari}, {Biland}, {Bisesi}, {Bock},
  {Bordas}, {Bosch-Ramon}, {Bretz}, {Britvitch}, {Camara}, {Carmona},
  {Chilingarian}, {Ciprini}, {Coarasa}, {Commichau}, {Contreras}, {Cortina},
  {Curtef}, {Danielyan}, {Dazzi}, {De Angelis}, {de los Reyes}, {De Lotto},
  {Domingo-Santamar{\'{\i}}a}, {Dorner}, {Doro}, {Errando}, {Fagiolini},
  {Ferenc}, {Fern{\'a}ndez}, {Firpo}, {Flix}, {Fonseca}, {Font}, {Fuchs},
  {Galante}, {Garczarczyk}, {Gaug}, {Giller}, {Goebel}, {Hakobyan},
  {Hayashida}, {Hengstebeck}, {H{\"o}hne}, {Hose}, {Hsu}, {Isar}, {Jacon},
  {Kalekin}, {Kosyra}, {Kranich}, {Laatiaoui}, {Laille}, {Lenisa}, {Liebing},
  {Lindfors}, {Lombardi}, {Longo}, {L{\'o}pez}, {L{\'o}pez}, {Lorenz},
  {Lucarelli}, {Majumdar}, {Maneva}, {Mannheim}, {Mansutti}, {Mariotti},
  {Mart{\'{\i}}nez}, {Mase}, {Mazin}, {Merck}, {Meucci}, {Meyer}, {Miranda},
  {Mirzoyan}, {Mizobuchi}, {Moralejo}, {Nilsson}, {O{\~n}a-Wilhelmi},
  {Ordu{\~n}a}, {Otte}, {Oya}, {Paneque}, {Paoletti}, {Paredes}, {Pasanen},
  {Pascoli}, {Pauss}, {Pavel}, {Pegna}, {Persic}, {Peruzzo}, {Piccioli},
  {Poller}, {Pooley}, {Prandini}, {Raymers}, {Rhode}, {Rib{\'o}}, {Rico},
  {Riegel}, {Rissi}, {Robert}, {Romero}, {R{\"u}gamer}, {Saggion},
  {S{\'a}nchez}, {Sartori}, {Scalzotto}, {Scapin}, {Schmitt}, {Schweizer},
  {Shayduk}, {Shinozaki}, {Shore}, {Sidro}, {Sillanp{\"a}{\"a}}, {Sobczynska},
  {Stamerra}, {Stark}, {Takalo}, {Temnikov}, {Tescaro}, {Teshima}, {Tonello},
  {Torres}, {Torres}, {Turini}, {Vankov}, {Vitale}, {Wagner}, {Wibig},
  {Wittek}, {Zanin}, \& {Zapatero}}]{2006Sci...312.1771A}
{Albert}, J., {Aliu}, E., {Anderhub}, H., {et~al.} 2006, Science, 312, 1771

\bibitem[{{Albert} {et~al.}(2008){Albert}, {Aliu}, {Anderhub}, {Antoranz},
  {Armada}, {Asensio}, {Baixeras}, {Barrio}, {Bartelt}, {Bartko}, {Bastieri},
  {Bavikadi}, {Bednarek}, {Berger}, {Bigongiari}, {Biland}, {Bisesi}, {Bock},
  {Bordas}, {Bosch-Ramon}, {Bretz}, {Britvitch}, {Camara}, {Carmona},
  {Chilingarian}, {Ciprini}, {Coarasa}, {Commichau}, {Contreras}, {Cortina},
  {Curtef}, {Danielyan}, {Dazzi}, {De Angelis}, {de los Reyes}, {De Lotto},
  {Domingo-Santamar{\'{\i}}a}, {Dorner}, {Doro}, {Errando}, {Fagiolini},
  {Ferenc}, {Fern{\'a}ndez}, {Firpo}, {Flix}, {Fonseca}, {Font}, {Fuchs},
  {Galante}, {Garczarczyk}, {Gaug}, {Giller}, {Goebel}, {Hakobyan},
  {Hayashida}, {Hengstebeck}, {H{\"o}hne}, {Hose}, {Hsu}, {Isar}, {Jacon},
  {Kalekin}, {Kosyra}, {Kranich}, {Laatiaoui}, {Laille}, {Lenisa}, {Liebing},
  {Lindfors}, {Lombardi}, {Longo}, {L{\'o}pez}, {L{\'o}pez}, {Lorenz},
  {Lucarelli}, {Majumdar}, {Maneva}, {Mannheim}, {Mansutti}, {Mariotti},
  {Mart{\'{\i}}nez}, {Mase}, {Mazin}, {Merck}, {Meucci}, {Meyer}, {Miranda},
  {Mirzoyan}, {Mizobuchi}, {Moralejo}, {Nilsson}, {O{\~n}a-Wilhelmi},
  {Ordu{\~n}a}, {Otte}, {Oya}, {Paneque}, {Paoletti}, {Paredes}, {Pasanen},
  {Pascoli}, {Pauss}, {Pavel}, {Pegna}, {Persic}, {Peruzzo}, {Piccioli},
  {Poller}, {Pooley}, {Prandini}, {Raymers}, {Rhode}, {Rib{\'o}}, {Rico},
  {Riegel}, {Rissi}, {Robert}, {Romero}, {R{\"u}gamer}, {Saggion},
  {S{\'a}nchez}, {Sartori}, {Scalzotto}, {Scapin}, {Schmitt}, {Schweizer},
  {Shayduk}, {Shinozaki}, {Shore}, {Sidro}, {Sillanp{\"a}{\"a}}, {Sobczynska},
  {Stamerra}, {Stark}, {Takalo}, {Temnikov}, {Tescaro}, {Teshima}, {Tonello},
  {Torres}, {Torres}, {Turini}, {Vankov}, {Vitale}, {Wagner}, {Wibig},
  {Wittek}, {Zanin}, \& {Zapatero}}]{2008arXiv0806.1865M}
{Albert}, J., {Aliu}, E., {Anderhub}, H., {et~al.} 2008, ArXiv e-prints,
  0806.1865

\bibitem[{{Ball} \& {Dodd}(2001)}]{2001PASA...18...98B}
{Ball}, L. \& {Dodd}, J. 2001, Publications of the Astronomical Society of
  Australia, 18, 98

\bibitem[{{Ball} \& {Kirk}(2000)}]{2000APh....12..335B}
{Ball}, L. \& {Kirk}, J.~G. 2000, Astroparticle Physics, 12, 335

\bibitem[{{Begelman} \& {Li}(1992)}]{1992ApJ...397..187B}
{Begelman}, M.~C. \& {Li}, Z.-Y. 1992, \apj, 397, 187

\bibitem[{{Blumenthal} \& {Gould}(1970)}]{1970RvMP...42..237B}
{Blumenthal}, G.~R. \& {Gould}, R.~J. 1970, Reviews of Modern Physics, 42, 237

\bibitem[{{Bogovalov} \& {Aharonian}(2000)}]{2000MNRAS.313..504B}
{Bogovalov}, S.~V. \& {Aharonian}, F.~A. 2000, \mnras, 313, 504

\bibitem[{{Bogovalov} \& {Khangoulyan}(2002)}]{2002AstL...28..373B}
{Bogovalov}, S.~V. \& {Khangoulyan}, D.~V. 2002, Astronomy Letters, 28, 373

\bibitem[{{Bogovalov} {et~al.}(2008){Bogovalov}, {Khangulyan}, {Koldoba},
  {Ustyugova}, \& {Aharonian}}]{2008MNRAS.tmp..570B}
{Bogovalov}, S.~V., {Khangulyan}, D.~V., {Koldoba}, A.~V., {Ustyugova}, G.~V.,
  \& {Aharonian}, F.~A. 2008, \mnras, 570

\bibitem[{{Bosch-Ramon} {et~al.}(2008){Bosch-Ramon}, {Khangulyan}, \&
  {Aharonian}}]{2008A&A...482..397B}
{Bosch-Ramon}, V., {Khangulyan}, D., \& {Aharonian}, F.~A. 2008, \aap, 482, 397

\bibitem[{{Casares} {et~al.}(2005{\natexlab{a}}){Casares}, {Ribas}, {Paredes},
  {Mart{\'{\i}}}, \& {Allende Prieto}}]{2005MNRAS.360.1105C}
{Casares}, J., {Ribas}, I., {Paredes}, J.~M., {Mart{\'{\i}}}, J., \& {Allende
  Prieto}, C. 2005{\natexlab{a}}, \mnras, 360, 1105

\bibitem[{{Casares} {et~al.}(2005{\natexlab{b}}){Casares}, {Rib{\'o}}, {Ribas},
  {Paredes}, {Mart{\'{\i}}}, \& {Herrero}}]{2005MNRAS.364..899C}
{Casares}, J., {Rib{\'o}}, M., {Ribas}, I., {et~al.} 2005{\natexlab{b}},
  \mnras, 364, 899

\bibitem[{{Chernyakova} \& {Illarionov}(1999)}]{1999MNRAS.304..359C}
{Chernyakova}, M.~A. \& {Illarionov}, A.~F. 1999, \mnras, 304, 359

\bibitem[{{Dubus}(2006{\natexlab{a}})}]{2006A&A...451....9D}
{Dubus}, G. 2006{\natexlab{a}}, \aap, 451, 9

\bibitem[{{Dubus}(2006{\natexlab{b}})}]{2006A&A...456..801D}
{Dubus}, G. 2006{\natexlab{b}}, \aap, 456, 801

\bibitem[{{Dubus} {et~al.}(2008){Dubus}, {Cerutti}, \&
  {Henri}}]{2008A&A...477..691D}
{Dubus}, G., {Cerutti}, B., \& {Henri}, G. 2008, \aap, 477, 691

\bibitem[{{Eichler} \& {Usov}(1993)}]{1993ApJ...402..271E}
{Eichler}, D. \& {Usov}, V. 1993, \apj, 402, 271

\bibitem[{{Fargion} {et~al.}(1997){Fargion}, {Konoplich}, \& {Salis}}]{fargion}
{Fargion}, D., {Konoplich}, R.~V., \& {Salis}, A. 1997, Z. Phys. C., 74, 571

\bibitem[{{Grundstrom} {et~al.}(2007){Grundstrom}, {Caballero-Nieves}, {Gies},
  {Huang}, {McSwain}, {Rafter}, {Riddle}, {Williams}, \&
  {Wingert}}]{2007ApJ...656..437G}
{Grundstrom}, E.~D., {Caballero-Nieves}, S.~M., {Gies}, D.~R., {et~al.} 2007,
  \apj, 656, 437

\bibitem[{{Hartman} {et~al.}(1999){Hartman}, {Bertsch}, {Bloom}, {Chen},
  {Deines-Jones}, {Esposito}, {Fichtel}, {Friedlander}, {Hunter}, {McDonald},
  {Sreekumar}, {Thompson}, {Jones}, {Lin}, {Michelson}, {Nolan}, {Tompkins},
  {Kanbach}, {Mayer-Hasselwander}, {M{\"u}cke}, {Pohl}, {Reimer}, {Kniffen},
  {Schneid}, {von Montigny}, {Mukherjee}, \& {Dingus}}]{1999ApJS..123...79H}
{Hartman}, R.~C., {Bertsch}, D.~L., {Bloom}, S.~D., {et~al.} 1999, \apjs, 123,
  79

\bibitem[{{Johnston} {et~al.}(1992){Johnston}, {Manchester}, {Lyne}, {Bailes},
  {Kaspi}, {Qiao}, \& {D'Amico}}]{1992ApJ...387L..37J}
{Johnston}, S., {Manchester}, R.~N., {Lyne}, A.~G., {et~al.} 1992, \apjl, 387,
  L37

\bibitem[{{Jones}(1965)}]{1965PhRv..137.1306J}
{Jones}, F.~C. 1965, Physical Review, 137, 1306

\bibitem[{{Kennel} \& {Coroniti}(1984)}]{1984ApJ...283..694K}
{Kennel}, C.~F. \& {Coroniti}, F.~V. 1984, \apj, 283, 694

\bibitem[{{Khangulyan} {et~al.}(2007){Khangulyan}, {Hnatic}, {Aharonian}, \&
  {Bogovalov}}]{2007MNRAS.380..320K}
{Khangulyan}, D., {Hnatic}, S., {Aharonian}, F., \& {Bogovalov}, S. 2007,
  \mnras, 380, 320

\bibitem[{{Kirk} {et~al.}(1999){Kirk}, {Ball}, \&
  {Skjaeraasen}}]{1999APh....10...31K}
{Kirk}, J.~G., {Ball}, L., \& {Skjaeraasen}, O. 1999, Astroparticle Physics,
  10, 31

\bibitem[{{Kirk} {et~al.}(2007){Kirk}, {Lyubarsky}, \&
  {Petri}}]{2007astro.ph..3116K}
{Kirk}, J.~G., {Lyubarsky}, Y., \& {Petri}, J. 2007, ArXiv Astrophysics
  e-prints, 0703.116

\bibitem[{{Komissarov} \& {Lyubarsky}(2004)}]{2004MNRAS.349..779K}
{Komissarov}, S.~S. \& {Lyubarsky}, Y.~E. 2004, \mnras, 349, 779

\bibitem[{{McSwain} {et~al.}(2004){McSwain}, {Gies}, {Huang}, {Wiita},
  {Wingert}, \& {Kaper}}]{2004ApJ...600..927M}
{McSwain}, M.~V., {Gies}, D.~R., {Huang}, W., {et~al.} 2004, \apj, 600, 927

\bibitem[{{Rees} \& {Gunn}(1974)}]{1974MNRAS.167....1R}
{Rees}, M.~J. \& {Gunn}, J.~E. 1974, \mnras, 167, 1

\bibitem[{{Romero} {et~al.}(2007){Romero}, {Okazaki}, {Orellana}, \&
  {Owocki}}]{2007A&A...474...15R}
{Romero}, G.~E., {Okazaki}, A.~T., {Orellana}, M., \& {Owocki}, S.~P. 2007,
  \aap, 474, 15

\bibitem[{{Sierpowska} \& {Bednarek}(2005)}]{2005MNRAS.356..711S}
{Sierpowska}, A. \& {Bednarek}, W. 2005, \mnras, 356, 711

\bibitem[{{Sierpowska-Bartosik} \& {Bednarek}(2008)}]{2008MNRAS.tmp..347S}
{Sierpowska-Bartosik}, A. \& {Bednarek}, W. 2008, \mnras, 347

\bibitem[{{Sierpowska-Bartosik} \& {Torres}(2008)}]{2008arXiv0801.3427S}
{Sierpowska-Bartosik}, A. \& {Torres}, D.~F. 2008, ArXiv e-prints, 0801.3427

\bibitem[{{Stevens} {et~al.}(1992){Stevens}, {Blondin}, \&
  {Pollock}}]{1992ApJ...386..265S}
{Stevens}, I.~R., {Blondin}, J.~M., \& {Pollock}, A.~M.~T. 1992, \apj, 386, 265

\bibitem[{{Waters} {et~al.}(1988){Waters}, {van den Heuvel}, {Taylor},
  {Habets}, \& {Persi}}]{1988A&A...198..200W}
{Waters}, L.~B.~F.~M., {van den Heuvel}, E.~P.~J., {Taylor}, A.~R., {Habets},
  G.~M.~H.~J., \& {Persi}, P. 1988, \aap, 198, 200

\bibitem[{{Zdziarski}(1989)}]{1989ApJ...342.1108Z}
{Zdziarski}, A.~A. 1989, \apj, 342, 1108

\bibitem[{{Zdziarski} {et~al.}(2008){Zdziarski}, {Neronov}, \&
  {Chernyakova}}]{2008arXiv0802.1174Z}
{Zdziarski}, A.~A., {Neronov}, A., \& {Chernyakova}, M. 2008, ArXiv e-prints,
  0802.1174

\end{thebibliography}

\end{document}